%% file: paper.tex
\documentclass[a4paper]{jpconf}

\input{_paper}

\begin{document}
\title{\LARGE \bf
Model predictive control of wakes for wind farm power tracking
}

\author{Arnold Sterle, Christian A. Hans, Jörg Raisch}

\address{This work was partially supported by the German Federal Ministry for Economic Affairs and Climate Action (BMWK), project No. 03EE2036C.\\
A. Sterle and J. Raisch are with Technische Universität Berlin, Control Systems Group, Germany. J. Raisch is also with Science of Intelligence, Research Cluster of Excellence.\\
C. A. Hans is with the University of Kassel, Automation and Sensorics in Networked Systems Group, Germany.}%
\ead{\{sterle, raisch\}@control.tu-berlin.de, hans@uni-kassel.de}

\begin{abstract}

In this paper, a \acl{mpc} scheme for wind farms is presented.
Our approach considers wake dynamics including their influence on local wind conditions and allows the tracking of a given power reference.
In detail, a Gaussian wake model is used in combination with observation points that carry wind condition information.
This allows the estimation of the rotor effective wind speeds at downstream turbines, based on which we deduce their power output.
Through different approximation methods, the associated finite horizon nonlinear optimization problem is reformulated in a \acl{miqcqp} fashion.
By solving the reformulated problem online, optimal yaw angles and axial induction factors are found.
Closed-loop simulations indicate good power tracking capabilities over a wide range of power setpoints while distributing \acl{wt} infeed evenly among all units.
Additionally, the simulation results underline real time capabilities of our approach.

\end{abstract}

\input{introduction}
\input{model}
\input{mpc}
\input{caseStudy}
\input{conclusions}

\section{Literature}
\bibliographystyle{IEEEtran}
\bibliography{literature}

\end{document}

%% file: _paper.tex
\usepackage[T1]{fontenc}
\usepackage[utf8]{inputenc}

\usepackage{amsmath}
\usepackage{amsthm}
\usepackage{nicefrac}
\usepackage{bm}
\usepackage{amsfonts}
\usepackage{gensymb}
\usepackage{leftindex}
\usepackage[per-mode=symbol]{siunitx}
\usepackage{cite}
\usepackage{booktabs}
\usepackage[capitalise]{cleveref}
\crefname{subsection}{Subsection}{Subsections}
\crefname{problem}{Problem}{Problems}
\crefname{assumption}{Ass.}{Ass.}
\theoremstyle{definition}
\newtheorem{problem}{Problem}
\newtheorem{assumption}{Assumption}
\usepackage{cases}
\usepackage{graphicx}
\usepackage{adjustbox}
\usepackage{subcaption}
\usepackage{cuted}
\usepackage{xcolor}
\definecolor{vgRed}{RGB}{193, 48, 24}
\definecolor{vgOrange}{RGB}{243, 111, 19}
\definecolor{vgYellow}{RGB}{235, 203, 56}
\definecolor{vgGreen}{RGB}{162, 185, 105}
\definecolor{vgLightBlue}{RGB}{13, 149, 188}
\definecolor{vgDarkBlue}{RGB}{6, 56, 81}

\usepackage[acronym,shortcuts]{glossaries}
\newacronym{bem}{BEM}{blade element momentum theory}
\newacronym{blc}{BLC}{baseline controller}
\newacronym{mpc}{MPC}{model predictive control}
\newacronym[longplural=wind turbines]{wt}{WT}{wind turbine}
\newacronym{pwa}{PWA}{piecewise affine}
\newacronym{mibp}{MIBP}{mixed integer bilinear program}
\newacronym{miqcqp}{MIQCQP}{mixed-integer quadratically-constrained quadratic program}
\newacronym{mimo}{MIMO}{multi-input-multi-output}
\newacronym[longplural=damage equivalent loads]{del}{DEL}{damage equivalent load}
\newacronym{rms}{RMS}{root mean square}
\newacronym[longplural=observation points]{op}{OP}{observation point}

\usepackage{tikz}
\usepackage{pgfplots}
\usetikzlibrary{arrows.meta,decorations.markings,calc,positioning,intersections,pgfplots.fillbetween}
\pgfplotsset{compat=1.15}
\usepgfplotslibrary{external}
\usepgfplotslibrary{statistics}
\tikzset{external/system call={pdflatex \tikzexternalcheckshellescape -halt-on-error
    -interaction=batchmode -jobname "\image" "\texsource"}}
\usepackage{todonotes}
\renewcommand{\todo}[2][]{\tikzexternaldisable\@todo[#1]{#2}\tikzexternalenable}

\pgfplotsset{every axis/.append style={semithick,tick style={major tick
            length=4pt,semithick,black}}}

\pgfplotsset{
    colormap={customColorMap}{
        rgb255=(6, 56, 81)
        rgb255=(13, 149, 188)
        rgb255=(162, 185, 105)
        rgb255=(235, 203, 56)
        rgb255=(243, 111, 19)
        rgb255=(193, 48, 24)
    },
}

\pgfplotsset{
    boxplot/whisker range = {100},
}

\pgfkeys{/pgfplots/x axis shift down/.style={
        x axis line style={yshift=-#1},
        xtick style={yshift=-#1},
        tick align=outside,
        xticklabel shift={#1}}}

\pgfkeys{/pgfplots/y axis shift left/.style={
        y axis line style={xshift=-#1},
        ytick style={xshift=-#1},
        yticklabel shift={#1},
        scaled y ticks = false,
        tick align=outside,
        y tick label style={/pgf/number format/fixed,
        }
    }
}

\pgfplotsset{myPlot/.style={%
        width=8cm,
        height=4cm,
        line width = 0.7pt,
        separate axis lines,
        axis x line*=bottom,
        x axis shift down = 3pt,
        enlarge x limits=false,
        axis y line*=left,
        y axis shift left = 6pt,
        enlarge y limits={abs=.25pt},
        enlarge x limits={abs=.25pt},
    }
}

\pgfdeclareshape{lowpassnode}
{
    \inheritsavedanchors[from={rectangle}]
    \inheritbackgroundpath[from={rectangle}]
    \inheritanchorborder[from={rectangle}]
    \foreach \x in {center,north east,north west,north,south,south east,south west, west, east}
    {
        \inheritanchor[from={rectangle}]{\x}
    }
        \foregroundpath
        {
            \pgfpointdiff{\northeast}{\southwest}
            \pgf@xa=\pgf@x \pgf@ya=\pgf@y
            \northeast
            \pgfpathmoveto{\pgfpoint{0.4\pgf@xa}{-0.2\pgf@ya}}
            \pgfpathlineto{\pgfpoint{0}{-0.2\pgf@ya}}
            \pgfpathlineto{\pgfpoint{-0.4\pgf@xa}{0.2\pgf@ya}}
            \pgfusepath{stroke}
        }
}

\newcommand{\Rp}{\mathbb{R}_{\geq0}}
\newcommand{\R}{\mathbb{R}}
\newcommand{\Rps}{\mathbb{R}_{>0}}

\newcommand{\N}{\mathbb{N}}

\newcommand{\xDelta}{x_\Delta}
\newcommand{\yDelta}{y_\Delta}
\DeclareMathOperator{\erf}{erf}
\newcommand{\erfpwa}{\erf^{PWA}}

%% file: introduction.tex

\section{Introduction}

In state-of-the-art wind farm control, actuation is computed based on open-loop control schemes which are typically deduced via offline optimizations using steady-state models \cite{Meyers_2022}.
Such approaches may find optimal steady-state solutions for constant wind conditions and power setpoints.
However, state and input trajectories driving the system from one state to another can lead to poor performance.
Closed-loop control concepts that are designed based on dynamic models bear the potential to improve the overall performance by finding more suitable trajectories under changing wind conditions and power setpoints.

In literature, a variety of closed-loop control approaches have been studied.
In \cite{Shapiro_2017}, a control scheme which employs partial differential equations to model wakes is investigated.
However, time dependencies on the thrust coefficient, which have critical influence on wake dynamics, are neglected.
Additionally, the wake model is based on the Jensen model \cite{Jensen1983}, which is typically less accurate than recent engineering models such as Gaussian wake models \cite{Bastankhah_2016} or cumulative curl models \cite{Bay_2022}.
In \cite{Vali2019}, a gain-scheduling proportional-integral controller for power tracking is studied.
While this approach reduces mechanical stress in the presence of changing power setpoints, its power tracking capabilities are limited by the induction control range of each turbine since yawing is not taken into account, even though it enables a larger range of available power \cite{Campagnolo_2020,Fleming_2020,Simley_2021}.
In \cite{Becker_2022,Becker_2022Revised}, the framework FLORIDyn is investigated.
FLORIDyn models wake dynamics by propagating wind conditions which can be influenced by turbines via \acfp{op}.
However, the approach is targeted towards simulation and cannot be directly used for control.

This paper presents a novel \ac{mpc} approach that uses a Gaussian wake model in tandem with the concepts from \cite{Becker_2022,Becker_2022Revised} to formulate a \ac{miqcqp}. The resulting controller is capable of finding state and input trajectories to track changing power setpoints by actively controlling wakes.
A proof-of-concept case study validates our approach which provides a broad basis for more complex \ac{mpc} schemes that consider, e.g., power tracking under heterogenous wind conditions.

The remainder of this paper is structured as follows.
In \cref{sec:model}, wake and turbine models are introduced.
Then, in \cref{sec:mpc}, these models are used to formulate an \ac{mpc} scheme.
In \cref{sec:caseStudy}, a case study is presented.
Finally, in \cref{sec:conclusion}, the results of this paper are discussed and an outlook on future work is given.

%% file: model.tex

\section{Model}\label{sec:model}

In this section, a control-oriented wind farm model composed of wake transport dynamics, Gaussian wake models and \ac{wt} model is presented.
Throughout this paper, the simple wind farm in \cref{fig:threeTurbineFarm} serves as an example.
However, the results can also be applied to more complex real-world setups.
We start with some general assumptions and remarks.
\begin{figure}[t]
    \centering
    \input{figures/threeTurbineFarm.tex}
    \vspace{-0.5cm}
    \caption{Simple example of a wind farm with three turbines and local coordinate systems.
    The solid circles represent \aclp{op} and the empty circles represent \aclp{wt}.}
	\label{fig:threeTurbineFarm}
\end{figure}
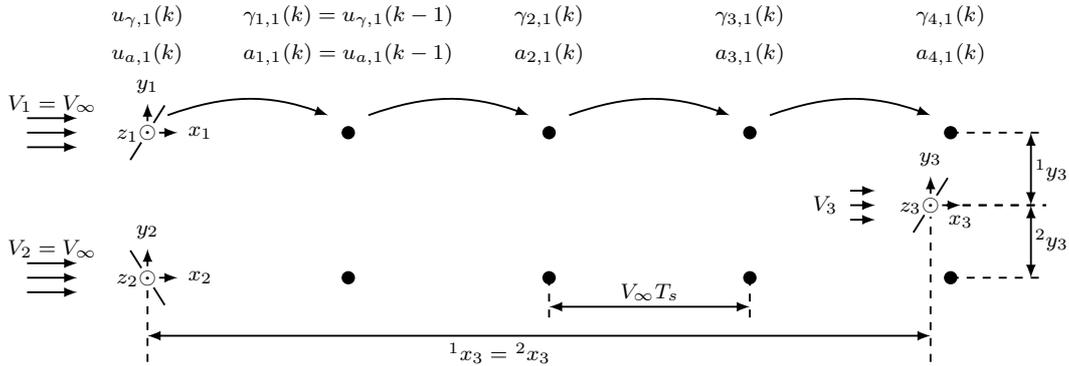
\begin{assumption}\label{ass:wind}
    We assume a persistence forecast of free wind conditions, i.e., within the finite prediction horizon of the controller, the free wind speed, $V_\infty\in\Rp$, is constant in magnitude and direction.
\end{assumption}
To describe spatial quantities, each \ac{wt} is equipped with a three-dimensional Cartesian coordinate system.
\Cref{ass:wind} allows us to align the $x$-axis of each coordinate system with the free wind direction.
Heights are represented by the $z$-coordinate, and the $y$-axis is perpendicular to the $x$- and $z$-axis.
We refer to $x,y,z$ directions as streamwise, spanwise and vertical, respectively.
The origin of each coordinate system, where $x=y=z=0$, is located at the centre of each \ac{wt}'s rotor.
\begin{assumption}
    For simplicity, all \acp{wt} are assumed to be identical such that they share equal parameters, e.g, diameters, hub height etc.
\end{assumption}
The following model aims to estimate the power output of each \ac{wt}, which depends on their rotor effective wind speeds.
As indicated in \cref{fig:threeTurbineFarm}, the rotor effective wind speed of each upstream \ac{wt} equals the free wind speed.
Since wind conditions at downstream \acp{wt} are influenced by wake, we need to model how actions of upstream \acp{wt} propagate through the wind field using transport dynamics.

\subsection{Transport Dynamics}\label{subsec:op}

The modelling of wake transport dynamics sketched in \cref{fig:threeTurbineFarm} is inspired by FLORIDyn \cite{Gebraad_2014_FLORIDyn,Becker_2022Revised,Becker_2022}.
Let $k\in\N_0$ denote the discrete time instant. 
Then, consider \ac{wt} $i$ with control inputs in the form of yaw misalignment $u_{\gamma,i}(k)\in\R$ and axial induction factor $u_{a,i}(k)\in\Rps$, which is short for $u_{\gamma,i}(kT_s)$ and $u_{a,i}(kT_s)$, where $T_s\in\Rps$ denotes the controller sampling time.
The axial induction factor is a measurement for how strongly a \ac{wt} decelerates the flow field (as long as it is below $0.4$, see, e.g., \cite{Hansen_2015}). Massless particles, called \acp{op}, are placed behind \ac{wt} $i$ in a distance of $V_\infty T_s$ in streamwise direction.
Each \ac{op} $m$ associated with \ac{wt} $i$ exhibits a yaw misalignment $\gamma_{m,i}(k)\in\R$ and an axial induction factor $a_{m,i}(k)\in\Rp$ which allow to approximate local wind conditions.
\begin{assumption}\label{ass:wakeVelocity}
    The local wind conditions are assumed to propagate with velocity $V_\infty$.
\end{assumption}
\Cref{ass:wakeVelocity} allows us to update each \ac{op} at each time step either with the values of its upstream \ac{op} or the values originating from the associated \ac{wt}.
Let $N_{OP}\in\N$ denote the number of \acp{op} per turbine.
Then, the values of the \acp{op} of \ac{wt} $i$, collected in ${\gamma_i=[\gamma_{1,i},\,\gamma_{2,i},\,\cdots\,\gamma_{N_{OP},i}]^T}$ and ${a_i=[a_{1,i},\,a_{2,i},\,\cdots\,a_{N_{OP},i}]^T}$ can be described by the dynamics
\begin{subequations}\label{eq:dynamics}
    \begin{align}
        \gamma_i(k+1) &= A \gamma_i(k) + B u_{\gamma,i}(k),\\
        a_i(k+1) &= A a_i(k) + B u_{a,i}(k).
    \end{align}
\end{subequations}
Here, $A\in\R^{N_{OP}\times N_{OP}}$ as well as $B\in\R^{N_{OP}}$ are given by

\vspace{0.5cm}
\hfill
$
    A = 
    \begin{bmatrix}
        0 & 0      & \cdots & 0\\
        1 & 0      & \cdots & 0\\
        & \ddots & \ddots & \vdots\\
        0 &        &      1 & 0
    \end{bmatrix}
$
\qquad and \qquad
$
    B =
    \begin{bmatrix}
        1 & 0 & \cdots & 0
    \end{bmatrix}^T.
$
\hfill
\vspace{0.5cm}

The \ac{op} closest to each downstream \ac{wt} will then be used to estimate local wind conditions for the respective \ac{wt} using the following wake model.

\subsection{Gaussian Wake Model}\label{subsec:gauss}
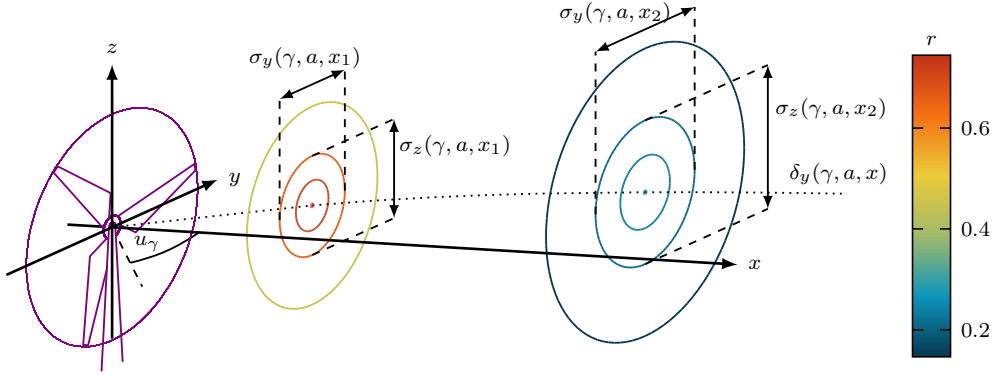
\begin{figure}[t]
    \centering
    \vspace{-0.1cm}
    \input{figures/3Dwake}
    \vspace{-0.5cm}
    \caption{
        Three dimensional wake forms and their defining parameters.
        From the turbine on the left, the wake form starts with smaller wake widths, $\sigma_y$ and $\sigma_z$, and higher wind speed deficit at the centre. 
        With increasing distance in $x$-direction, the wake widths become larger, which leads to a flatter wind speed deficit distribution.
        Additionally, the centre wind speed deficit decreases.
        Due to the nonzero yaw misalignment, $u_\gamma$, of the \ac{wt} the wake centre is deflected by $\delta_y$.
    }
    \label{fig:3Dwake}
\end{figure}

In this section, the Gaussian wake model from \cite{Bastankhah_2016}, which is used to estimate local wind conditions, will be recalled.
The defining parameters are sketched in \cref{fig:3Dwake}.

For simplicity, subscripts for the yaw misalignment and axial induction factor are omitted in this section.
Thus, consider an \ac{op} with yaw misalignment $\gamma$ and axial induction factor $a$.
Then, its associated thrust coefficient $C_t\in\Rps$ is
\begin{equation}\label{eq:thrustCoeffcient}
    C_t(\gamma,a) = 4a\left(1-a\cos(\gamma)\right).
\end{equation}
The wake core length $x_c\in\Rp$ is
\begin{equation}\label{eq:x0}
    x_c(\gamma,a) = \textstyle D\frac{\cos(\gamma)\left(1+\sqrt{1-C_t}\right)}{\sqrt{2}\left(\alpha I+\beta\left(1-\sqrt{1-C_t}\right)\right)},
\end{equation}
where ${D\in\Rp}$ denotes the \ac{wt}'s diameter and and ${\alpha,\beta\in\Rp}$ are constants which describe the wake recovery rate.
The wake core length marks the distance in downstream direction that separates near wake conditions ($x<x_c$) from far wake conditions for ($x\geq x_c$).
This is important since different models are used to describe the flow field under these different conditions.
\begin{assumption}\label{ass:turbulence}
    The turbulence intensity, $I\in\Rp$, is constant.
\end{assumption}
\begin{assumption}\label{ass:distance}
    Streamwise distances between turbines are much larger than their diameters, such that far wake conditions can be considered at all downstream turbines.
    Consequently, we can assume ${x\geq x_c}$.
\end{assumption}

At far wake conditions, the local wind speed deficit $r\in\Rp$ can be described by \cite{Bastankhah_2016}
\begin{equation}\label{eq:localWindDeficit}
    r(\gamma,a,x,y,z) = \textstyle r_c(\gamma,a,x)\exp\left(-\left(\frac{y-\delta_y(\gamma,a,x)}{\sqrt{2}\sigma_y(\gamma,a,x)}\right)^2\right)\exp\left(-\left(\frac{z}{\sqrt{2}\sigma_z(\gamma,a,x)}\right)^2\right).
\end{equation}
The spanwise and vertical wake widths, $\sigma_y,\sigma_z\in\Rp$, are given by
\begin{subequations}\label{eq:sigma}
    \begin{align}
        \sigma_y(\gamma,a,x) &= \textstyle k_y(x-x_c(\gamma,a))+\frac{D}{\sqrt{8}}\cos(\gamma),\label{eq:sigmay}\\
        \sigma_z(\gamma,a,x) &= k_z(x-x_c(\gamma,a))+\textstyle\frac{D}{\sqrt{8}},\label{eq:sigmaz}
    \end{align}
\end{subequations}
where $k_y=k_z\in\Rp$ describe spanwise and vertical wake growth rates which linearly depend on the turbulence intensity.
Moreover, the wind speed deficit at the wake centre is
\begin{equation}\label{eq:C}
    r_c(\gamma,a,x) = \textstyle1-\sqrt{1-\frac{D^2C_t(\gamma,a)\cos(\gamma)}{8\sigma_y(\gamma,a,x)\sigma_z(\gamma,a,x)}}.
\end{equation}
Finally, the wake centre's spanwise deflection is given by
\begin{multline}\label{eq:deltay}
    \delta_y(\gamma,a,x) = \textstyle\zeta(\gamma,a) x_c+D\frac{\zeta(\gamma,a)}{14.7}\sqrt{\frac{cos(\gamma)}{k_yk_zC_t(\gamma,a)}}\left(2.9+1.3\sqrt{1-C_t(\gamma,a)}-C_t(\gamma,a)\right)\\
    \textstyle\ln\left(\frac{\left(1.6+\sqrt{C_t(\gamma,a)}\right)\left(1.6\sqrt{\frac{8\sigma_y(\gamma,a,x)\sigma_z(\gamma,a,x)}{D^2\cos(\gamma)}}-\sqrt{C_t(\gamma,a)}\right)}{\left(1.6-\sqrt{C_t(\gamma,a)}\right)\left(1.6\sqrt{\frac{8\sigma_y(\gamma,a,x)\sigma_z(\gamma,a,x)}{D^2\cos(\gamma)}}+\sqrt{C_t(\gamma,a)}\right)}\right),
\end{multline}
with skew angle
\begin{equation}\label{eq:zeta}
    \zeta(\gamma,a) = \textstyle\frac{0.3\gamma}{\cos(\gamma)}\left(1-\sqrt{1-C_t(\gamma,a)\cos(\gamma)}\right).
\end{equation}
Eqs. \labelcref{eq:thrustCoeffcient,eq:x0,eq:localWindDeficit,eq:sigma,eq:C,eq:deltay,eq:zeta} can now be employed to calculate local wind conditions at downstream \acp{wt}.

\subsection{Wind Turbine Model}

\begin{assumption}\label{ass:dynamics}
    We expect the turbine dynamics to be significantly faster than the farm dynamics described in \eqref{eq:dynamics}. Therefore, the \ac{wt} dynamics can be neglected.
    However, the yaw misalignment's rate of change is assumed to be limited.   
\end{assumption}
In the following, we focus on estimating generated power for given local wind conditions.
Since each downstream turbine is influenced by wakes we first require the rotor effective wind speed which we will deduce via the local wind speed deficit \eqref{eq:localWindDeficit}.
Consider an \ac{op} with yaw-misalignment $\gamma$ and axial induction factor $a$ which is associated with \ac{wt} $i$.
This \ac{op} is located close to a downstream \ac{wt} $j$ with yaw misalignment $u_{\gamma,j}$ which has the $i$-coordinates $(\leftindex^i  x_{j},\leftindex^i y_{j},\leftindex^i z_{j})$ (compare \cref{fig:threeTurbineFarm}).
The local wind speed deficit that originates from \ac{wt} $i$ is described by $r_i(\gamma,a,\leftindex^i  x_{j},\leftindex^i y_{j},\leftindex^i z_{j})$.
To compute the rotor effective wind speed deficit $R_{i,j}(\gamma,a,\leftindex^i x_{j},\leftindex^i y_{j},u_{\gamma,j})$, we need to integrate $r_i(\gamma,a,\leftindex^i x_{j},\leftindex^i y_{j},\leftindex^i z_{j})$ along \ac{wt} $j$'s rotor surface and divide the result by its surface area (see \cref{fig:rIntegral}).
\begin{figure}[b]
    \centering
    \vspace{-0.5cm}
    \input{figures/3DwakeIntegral}
    \vspace{-0.7cm}
    \caption{Point on downstream \ac{wt} rotor disc and rectangular approximation.}
    \label{fig:rIntegral}
\end{figure}
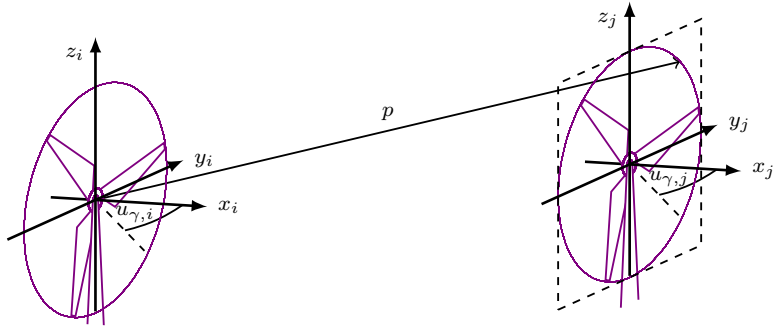
\begin{assumption}\label{ass:yaw}
    The yaw misalignment is limited to $u_{\gamma,i}\in[-\nicefrac{\pi}{6},\nicefrac{\pi}{6}]$ (see, e.g., \cite{Jim_nez_2009}).
\end{assumption}
We can describe any point $p$ in the surface area of \ac{wt} $j$'s rotor disc with ${l\in[0,\nicefrac{D}{2}]}$ and ${\phi\in[0,2\pi)}$.
Then, with \cref{ass:yaw}, we can deduce ${|l\sin(u_{\gamma,j})\cos(\phi)|\leq|\nicefrac{D}{2}\sin\left(\nicefrac{\pi}{6}\right)\cos(0)|=\nicefrac{D}{4}}$.
With ${\nicefrac{D}{4}\ll \leftindex^i x_{j}}$ (recall \cref{ass:distance}), the following approximation is justified:
\begin{equation}\label{eq:p}
    p=
    \begin{bmatrix}
        \leftindex^i x_{j} + l\sin(u_{\gamma,j})\cos(\phi)\\
        \leftindex^i y_{j} + l\cos(u_{\gamma,j})\cos(\phi)\\
        l\sin(\phi)
    \end{bmatrix}
    \approx
    \begin{bmatrix}
        \leftindex^i x_{j}\\
        \leftindex^i y_{j} + l\cos(u_{\gamma,j})\cos(\phi)\\
        l\sin(\phi)
    \end{bmatrix},
\end{equation}
To compute $R_{i,j}(\gamma,a,\leftindex^i x_{j},\leftindex^i y_{j},u_{\gamma,j})$ by integrating $r_i(\gamma,a,\leftindex^i x_{j},\leftindex^i y_{j},\leftindex^i z_{j})$ over $l$ and $\phi$, the Jacobian determinant is needed.
Since the $x$-component is considered constant in \eqref{eq:p}, the Jacobian determinant is given by
\begin{equation}
    \textstyle \det\left(\frac{\partial(p_y,p_z)}{\partial(l,\phi)}\right) = \textstyle\det\left(
        \begin{bmatrix}
            \cos(u_{\gamma,j})\cos(\phi) & -l\cos(u_{\gamma,j})\sin(\phi)\\
            \sin(\phi) & l\cos(\phi)
        \end{bmatrix}
    \right) = l\cos(u_{\gamma,j}).
\end{equation}
With this, we can deduce 
\begin{subequations}
    \begin{align}
        R_{i,j}(\gamma,a,\leftindex^i x_{j},\leftindex^i y_{j},u_{\gamma,j})
        &=\textstyle \frac{\int_{0}^{2\pi}
        \int_{0}^{\frac{D}{2}}
        r_i\left(\gamma,a,\leftindex^i x_{j},\leftindex^i y_{j}+l\cos(u_{\gamma,j})\cos(\phi),l\sin(\phi)\right)l\cos(u_{\gamma,j})\,\mathrm{d}l\,\mathrm{d}\phi}{\int_{0}^{2\pi}
        \int_{0}^{\frac{D}{2}}l\cos(u_{\gamma,j})\,\mathrm{d}l\,\mathrm{d}\phi}\\
        &=\textstyle \frac{4}{\pi D^2}\int_{0}^{2\pi}
        \int_{0}^{\frac{D}{2}}
        r_i\left(\gamma,a,\leftindex^i x_{j},\leftindex^i y_{j}+l\cos(u_{\gamma,j})\cos(\phi),l\sin(\phi)\right)l\,\mathrm{d}l\,\mathrm{d}\phi.
    \end{align}
\end{subequations}
This integral, however, is hard to solve analytically. 
Therefore, we approximate it by using a $D\times D\cos(u_{\gamma,j})$ rectangle, instead (see \cref{fig:rIntegral}), i.e.,
\begin{subequations}\label{eq:rInt}
    \begin{equation}\label{eq:rIntInt}
        R_{i,j}(\gamma,a,\leftindex^i x_{j},\leftindex^i y_{j},u_{\gamma,j}) \approx\textstyle
        \frac{1}{D^2\cos(u_{\gamma,j})}
        \int_{-\frac{1}{2}D}^{\frac{1}{2}D}
        \int_{\leftindex^i y_{j}-\frac{1}{2}D\cos(u_{\gamma,j})}^{\leftindex^i y_{j}+\frac{1}{2}D\cos(u_{\gamma,j})}
        r_i(\gamma,a,\leftindex^i x_{j},y,z)\,\mathrm{d}y\,\mathrm{d}z,\hspace{4mm}
    \end{equation}
    \begin{equation}\label{eq:rIntSol}
        \hspace{7mm} = \textstyle \frac{\pi r_C\sigma_y\sigma_z}{D^2\cos(u_{\gamma,j})}
        \left(\erf\left(\textstyle \frac{\leftindex^i y_{j}+\frac{1}{2}D\cos(u_{\gamma,j})-\delta_y}{\sqrt{2}\sigma_y}\right)
        -\erf\left(\frac{\leftindex^i y_{j}-\frac{1}{2}D\cos(u_{\gamma,j})-\delta_y}{\sqrt{2}\sigma_y}\right)\right)
        \erf\left({\textstyle \frac{{1}}{\sqrt{8}}}\frac{D}{\sigma_z}\right),
    \end{equation}
\end{subequations}
where $\erf(t)$ denotes the error function which is defined as ${\frac{2}{\sqrt{\pi}}\int_{0}^{t}\exp(-\tau^2)\,\mathrm{d}\tau}$.

Let the set ${\mathbb{U}_{j}\subset\N}$ contain all \acp{wt} $i$ that affect \ac{wt} $j$.
Motivated by \cite{Becker_2022Revised}, the rotor effective wind speed $V_{j}$ at \ac{wt} $j$ is calculated as
\begin{equation}\label{eq:rotorEffectiveWindSpeed}
    V_{j} = V_\infty\textstyle\prod_{i\in\mathbb{U}_{j}}(1-R_{i,j}).
\end{equation}

Now that we have the rotor effective wind speeds, we can deduce the power output of each \ac{wt} $i$ via
\begin{equation}\label{eq:turbinePower}
    P_i = \textstyle\frac{\eta\rho\pi D^2}{8}V_i^3C_{p,i}(u_{a,i})\cos(u_{\gamma,i})^{p_p},
\end{equation}
where $\eta\in\Rp$ denotes the \ac{wt}'s efficiency, $\rho\in\Rp$ the air density and $,p_p\in\Rp$ the power-yaw loss exponent \cite{Hansen_2015,Liew2020}.
In literature, various numbers for $p_p$ have been reported. 
For instance, \cite{Campagnolo2020} reports $p_p=2.1$, \cite{Krogstad2011,Bartl2018} $p_p=3$ and \cite{Fleming2014} $p_p=1.8$.
As a middle ground, in this work, $p_p=2$ is chosen.
The power coefficient $C_{p,i}\in\R$ \cite{Hansen_2015} is given by
\begin{equation}\label{eq:powerCoefficient}
    C_{p,i}(u_{a,i}) = 4\kappa u_{a,j}\left(1-u_{a,i}\right)^2,
\end{equation}
where $\kappa\in\R$ is a correction factor to account for the fact that real rotor discs are not ideal.
Finally, the power generated by the farm can be deduced by summing over all $N_T$ \acp{wt}, i.e.,
\begin{equation}\label{eq:farmPower}
    P_F=\textstyle\sum_{i=1}^{N_T} P_i.
\end{equation}

%% file: figures/threeTurbineFarm.tex

\newcommand{\yTone}{4.75}
\newcommand{\yTtwo}{2.25}
\newcommand{\yTthree}{3.5}
\tikzset{external/export next=false, >=latex}
  \begin{tikzpicture}[font=\scriptsize]
    \begin{axis}[
      myPlot,
      height = 5cm,
      width = 15cm,
      clip mode = individual,
      x axis line style = {draw = none},
      x tick style = {draw = none},
      xticklabels = {},
      y axis line style = {draw = none},
      y tick style = {draw = none},
      yticklabels = {},
      ]
      

    \node [circle, fill = black, minimum size = 5pt, inner sep = 0pt] at (2,\yTone) {};
    \node [circle, fill = black, minimum size = 5pt, inner sep = 0pt] at (4,\yTone) {};
    \node [circle, fill = black, minimum size = 5pt, inner sep = 0pt] at (6,\yTone) {};
    \node [circle, fill = black, minimum size = 5pt, inner sep = 0pt] at (8,\yTone) {};

    \node [circle, fill = black, minimum size = 5pt, inner sep = 0pt] at (2,\yTtwo) {};
    \node [circle, fill = black, minimum size = 5pt, inner sep = 0pt] at (4,\yTtwo) {};
    \node [circle, fill = black, minimum size = 5pt, inner sep = 0pt] at (6,\yTtwo) {};
    \node [circle, fill = black, minimum size = 5pt, inner sep = 0pt] at (8,\yTtwo) {};

    \addplot [->] coordinates {
      (-1.2,\yTone+0.25)
      (-0.7,\yTone+0.25)
    };
    \addplot [->] coordinates {
      (-1.2,\yTone)
      (-0.7,\yTone)
    };
    \addplot [->] coordinates {
      (-1.2,\yTone-0.25)
      (-0.7,\yTone-0.25)
    };
    \node at (-0.95,\yTone+0.5) {$V_1=V_\infty$};

    \addplot [->] coordinates {
      (-1.2,\yTtwo+0.25)
      (-0.7,\yTtwo+0.25)
    };
    \addplot [->] coordinates {
      (-1.2,\yTtwo)
      (-0.7,\yTtwo)
    };
    \addplot [->] coordinates {
      (-1.2,\yTtwo-0.25)
      (-0.7,\yTtwo-0.25)
    };
    \node at (-0.95,\yTtwo+0.5) {$V_2=V_\infty$};

    \addplot [->] coordinates {
      (8-1.0,\yTthree+0.25)
      (8-0.75,\yTthree+0.25)
    };
    \addplot [->] coordinates {
      (8-1.0,\yTthree)
      (8-0.75,\yTthree)
    };
    \addplot [->] coordinates {
      (8-1.0,\yTthree-0.25)
      (8-0.75,\yTthree-0.25)
    };
    \node at (8-1.25,\yTthree) {$V_3$};

    \node (ug1) at (0,\yTone+2) {$u_{\gamma,1}(k)$};
    \node (ua1) [below=0.5cm of ug1.west, anchor = west] {$u_{a,1}(k)$};

    \node (g11) at (2,\yTone+2) {$\gamma_{1,1}(k) = u_{\gamma,1}(k-1)$};
    \node (a11) [below=0.5cm of g11.west, anchor = west] {$a_{1,1}(k) = u_{a,1}(k-1)$};

    \node (g21) at (4,\yTone+2) {$\gamma_{2,1}(k)$};
    \node [below=0.5cm of g21.west, anchor = west] {$a_{2,1}(k)$};

    \node (g31) at (6,\yTone+2) {$\gamma_{3,1}(k)$};
    \node [below=0.5cm of g31.west, anchor = west] {$a_{3,1}(k)$};

    \node (g41) at (8,\yTone+2) {$\gamma_{4,1}(k)$};
    \node [below=0.5cm of g41.west, anchor = west] {$a_{4,1}(k)$};

    \path [->] (0+0.2,\yTone+0.3) edge [bend left = 20] node [above] {} (2-0.2,\yTone+0.3);
    \path [->] (2+0.2,\yTone+0.3) edge [bend left = 20] node [above] {} (4-0.2,\yTone+0.3);
    \path [->] (4+0.2,\yTone+0.3) edge [bend left = 20] node [above] {} (6-0.2,\yTone+0.3);
    \path [->] (6+0.2,\yTone+0.3) edge [bend left = 20] node [above] {} (8-0.2,\yTone+0.3);

    \addplot [<->] coordinates {
      (4,\yTtwo-0.5)
      (6,\yTtwo-0.5)
    };
    \node at (5,\yTtwo-0.25) {$V_\infty T_s$};
    \addplot [dashed] coordinates {
      (4,\yTtwo)
      (4,\yTtwo-0.75)
    };
    \addplot [dashed] coordinates {
      (6,\yTtwo)
      (6,\yTtwo-0.75)
    };

    \addplot [<->] coordinates {
      (7.8+1,\yTone)
      (7.8+1,\yTthree)
    };
    \node at (8+1,4.1) {$\leftindex^1 y_3$};
    \addplot [dashed] coordinates {
      (8,\yTone)
      (8+0.95,\yTone)
    };
    \addplot [dashed] coordinates {
      (7.8,\yTthree)
      (8+0.95,\yTthree)
    };
    \addplot [<->] coordinates {
      (7.8+1,\yTtwo)
      (7.8+1,\yTthree)
    };
    \node at (8+1,2.9) {$\leftindex^2 y_3$};
    \addplot [dashed] coordinates {
      (8,\yTtwo)
      (8+0.95,\yTtwo)
    };
    \addplot [dashed] coordinates {
      (7.8,\yTthree)
      (8+0.95,\yTthree)
    };

    \addplot [<->] coordinates {
      (0,\yTtwo-1)
      (7.8,\yTtwo-1)
    };
    \node at (3.5,\yTtwo-1.3) {$\leftindex^1 x_3 = \leftindex^2 x_3$};
    \addplot [dashed] coordinates {
      (0,0.8)
      (0,\yTtwo)
    };
    \addplot [dashed] coordinates {
      (7.8,0.8)
      (7.8,\yTthree)
    };

    \addplot[black] coordinates {
      (0-0.171,\yTone -0.4698)
      (0+0.171,\yTone +0.4698)
    };
    \node [circle, fill=black, minimum size = 7pt, inner sep = 0pt] at (0,\yTone) {};

    \addplot[black] coordinates {
      (0-0.171,\yTtwo+0.4698)
      (0+0.171,\yTtwo-0.4698)
    };
    \node [circle, fill=black, minimum size = 7pt, inner sep = 0pt] at (0,\yTtwo) {};

    \addplot[black] coordinates {
      (7.8+0.171,\yTthree+0.4698)
      (7.8-0.171,\yTthree-0.4698)
    };
    \node [circle, fill=black, minimum size = 7pt, inner sep = 0pt] at (7.8,\yTthree) {};

    \draw[->] (0,\yTone) -- (0.3,\yTone) node [right] {$x_1$};
    \draw[->] (0,\yTone) -- (0,\yTone+0.5) node [above] {$y_1$};
    \node[circle, fill=white, minimum size = 4pt, inner sep = 0pt] at (0,\yTone) {$\odot$};
    \node at (-0.2,4.7) {$z_1$};

    \draw[->] (0,\yTtwo) -- (0.3,\yTtwo) node [right] {$x_2$};
    \draw[->] (0,\yTtwo) -- (0,\yTtwo+0.5) node [above] {$y_2$};
    \node[circle, fill=white, minimum size = 4pt, inner sep = 0pt] at (0,\yTtwo) {$\odot$};
    \node at (-0.2,2.2) {$z_2$};

    \draw[->] (7.8,\yTthree) -- (7.8+0.3,\yTthree) node [right,below] {$x_3$};
    \draw[->] (7.8,\yTthree) -- (7.8,\yTthree+0.5) node [above] {$y_3$};
    \node[circle, fill=white, minimum size = 4pt, inner sep = 0pt] at (7.8,\yTthree) {$\odot$};
    \node at (7.8-0.2,3.45) {$z_3$};
  \end{axis}
\end{tikzpicture}

%% file: figures/3Dwake.tex

\newcommand{\h}{0.9154}
\tikzset{external/export next=false}
\begin{tikzpicture}[font=\scriptsize, >=latex]
    \begin{axis}[
        width = 12cm,
        height = 7cm,
        line width = 0.7pt,
        x axis line style = {draw = none},
        x tick style = {draw = none},
        xticklabels = {},
        y axis line style = {draw = none},
        y tick style = {draw = none},
        yticklabels = {},
        z axis line style = {draw = none},
        z tick style = {draw = none},
        zticklabels = {},
        xmin = -2,
        xmax = 11,
        ymin = -0.1,
        ymax = 0.9,
        zmin = 0.5,
        zmax = 1.5,
        view={20}{20},
        colorbar,
        colorbar style = {
            ylabel = $r$,
            ylabel style = {rotate = -90, at={(0.5, 1.0)}, anchor=south},
            height=4cm,
            at={(1.15, 0.8)},
            },
        clip=false,
        ]

        \draw[dotted] (axis cs: 0, 0, \h) to (axis cs: 1.9643, 0.2057, \h) to (axis cs: 3, 0.3127, \h);

        \addplot3 [mesh, point meta=explicit] table [x expr=3.0, y=y, z=z, meta=h, col sep=comma]
            {figures/csvFiles/wake1ring1.csv};
        
        \addplot3 [mesh, point meta=explicit] table [x expr=3.0, y=y, z=z, meta=h, col sep=comma]
            {figures/csvFiles/wake1ring2.csv};
        
        \addplot3 [mesh, point meta=explicit] table [x expr=3.0, y=y, z=z, meta=h, col sep=comma]
            {figures/csvFiles/wake1ring3.csv};
        
        \draw[dotted] (axis cs: 3, 0.3127, \h) to (axis cs: 4, 0.3910, \h) to (axis cs: 5, 0.4538, \h) to (axis cs: 6, 0.5057, \h) to (axis cs: 7, 0.5494, \h) to (axis cs: 9, 0.6194, \h);
        \addplot3 [mesh, point meta=explicit] table [x expr=3.0, y=y, z=z, meta=h, col sep=comma]
            {figures/csvFiles/wake1ring0.csv};

        \addplot3 [mesh, point meta=explicit] table [x expr=9.0, y=y, z=z, meta=h,col sep=comma]
            {figures/csvFiles/wake3ring1.csv};
        
        \addplot3 [mesh, point meta=explicit] table [x expr=9.0, y=y, z=z, meta=h,col sep=comma]
            {figures/csvFiles/wake3ring2.csv};
        
        \addplot3 [mesh, point meta=explicit] table [x expr=9.0, y=y, z=z, meta=h,col sep=comma]
            {figures/csvFiles/wake3ring3.csv};

        \draw[line width = 1pt] (axis cs: -1, 0, \h) to  (axis cs: 0, 0, \h);

        \draw[dotted] (axis cs: 9, 0.6194, \h) to (axis cs: 10, 0.6479, \h) to (axis cs: 11, 0.6732, \h) to (axis cs: 12, 0.6957, \h) to (axis cs: 13, 0.7160, \h);
        \addplot3 [mesh, point meta=explicit] table [x expr=9.0, y=y, z=z, meta=h, col sep=comma]
        {figures/csvFiles/wake3ring0.csv};

        \node at (axis cs: 12, 0.9, \h) {$\delta_y(\gamma,a,x)$};

        \draw[dashed] (axis cs: 3, 0.3127+0.3086/2, \h) to (axis cs: 3, 0.3127+0.3086/2, 1.35);
        \draw[dashed] (axis cs: 3, 0.3127-0.3086/2, \h) to (axis cs: 3, 0.3127-0.3086/2, 1.35);
        \draw[<->] (axis cs: 3, 0.3127+0.3086/2, 1.35) to (axis cs: 3, 0.3127-0.3086/2, 1.35);
        \node at (axis cs: 2.9, 0.3, 1.45) {$\sigma_y(\gamma, a, x_1)$};

        \draw[dashed] (axis cs: 9, 0.6194+0.4688/2, \h) to (axis cs: 9, 0.6194+0.4688/2, 1.5);
        \draw[dashed] (axis cs: 9, 0.6194-0.4688/2, \h) to (axis cs: 9, 0.6194-0.4688/2, 1.5);
        \draw[<->] (axis cs: 9, 0.6194+0.4688/2, 1.5) to (axis cs: 9, 0.6194-0.4688/2, 1.5);
        \node at (axis cs: 8.9, 0.5, 1.6) {$\sigma_y(\gamma, a, x_2)$};

        \draw[dashed] (axis cs: 3, 0.3127, \h+0.3559/2) to (axis cs: 3, 0.7, \h+0.3559/2);
        \draw[dashed] (axis cs: 3, 0.3127, \h-0.3559/2) to (axis cs: 3, 0.7, \h-0.3559/2);
        \draw[<->] (axis cs: 3, 0.7, \h+0.3559/2) to (axis cs: 3, 0.7, \h-0.3559/2);
        \node at (axis cs: 2, 1.2, \h-0.1) {$\sigma_z(\gamma, a, x_1)$};

        \draw[dashed] (axis cs: 9, 0.6194, \h+0.5161/2) to (axis cs: 9, 1.2, \h+0.5161/2);
        \draw[dashed] (axis cs: 9, 0.6194, \h-0.5161/2) to (axis cs: 9, 1.2, \h-0.5161/2);
        \draw[<->] (axis cs: 9, 1.2, \h+0.5161/2) to (axis cs: 9, 1.2, \h-0.5161/2);
        \node at (axis cs: 9, 1.5, \h) {$\sigma_z(\gamma, a, x_2)$};
   
        \draw[dashed] (axis cs: 0,0,\h) to (axis cs: 3.2,-0.5334,\h);
        \path (axis cs: 1.92,-0.3192,\h) edge [bend right = 10] (axis cs: 1.9464,0,\h);
        \node at (axis cs: 1.4,-0.13,\h) {$u_{\gamma}$};
       
        \path [draw=violet, fill=none] 
        (axis cs: 0, -0.0043412, \h+0.02462019) --
        (axis cs: 0, 0.3895819, \h+0.09407946) --
        (axis cs: 0, 0.3939231,\h+0.06945927) --
        (axis cs: 0, 0.11150439, \h-0.05649576) --
        cycle;
    \path [draw=violet, fill=none] 
        (axis cs: 0, -0.01915111, \h-0.01606969) --
        (axis cs: 0, -0.27626615, \h+0.29034809) --
        (axis cs: 0, -0.25711504,\h+0.30641778) --
        (axis cs: 0, -0.00682543, \h+0.12481352) -- cycle;
    \path [draw=violet, fill=none] 
        (axis cs: 0, 0.02349232, \h-0.0085505) --
        (axis cs: 0, -0.11331574, \h-0.38442755) --
        (axis cs: 0, -0.13680806 ,\h -0.37587705) --
        (axis cs: 0, -0.10467896, \h-0.06831775) -- cycle;
    \addplot3[draw=violet, domain=0:360, variable=t, smooth, fill=white] (0,{0+0.04*cos(t)},{\h+0.04*sin(t)}) -- cycle;
    \addplot3[draw=violet, domain=0:360, variable=t, smooth, fill=white] (0,{0+0.015*cos(t)},{\h+0.015*sin(t)}) -- cycle; 
    \path [draw=none, fill=white]
        (axis cs: 0, 0.04,\h-0.6)  --
        (axis cs: 0, .015,\h) --
        (axis cs: 0, -.015,\h) --
        (axis cs: 0, -0.04,\h-0.6);
    \path [draw=violet, fill=white] 
        (axis cs: 0, 0.05,\h-0.5)  --
        (axis cs: 0, .015,\h)
        (axis cs: 0, -.015,\h) --
        (axis cs: 0, -0.05,\h-0.5);
    \addplot3[draw=violet, thin, domain=0:360, variable=t, smooth, fill=none] (0,{0+0.406*cos(t)},{\h+0.406*sin(t)}) -- cycle;
 
    \draw[line width = 1pt, ->] (axis cs: 0, 0, \h) to  (axis cs: 14, 0, \h) node [right] {$x$};
    \draw[line width = 1pt, ->] (axis cs: 0, -0.5, \h) to  (axis cs: 0, 0.5, \h) node [right] {$y$};
    \draw[line width = 1pt, ->] (axis cs: 0, 0, \h-0.4) to (axis cs: 0, 0, 1.5) node [above] {$z$};
    \end{axis}
\end{tikzpicture}

%% file: figures/3DwakeIntegral.tex

\newcommand{\h}{0.9154}
\tikzset{external/export next=false}
\begin{tikzpicture}[font=\scriptsize, >=latex]
    \begin{axis}[
        width = 12cm,
        height = 7cm,
        line width = 0.7pt,
        x axis line style = {draw = none},
        x tick style = {draw = none},
        xticklabels = {},
        y axis line style = {draw = none},
        y tick style = {draw = none},
        yticklabels = {},
        z axis line style = {draw = none},
        z tick style = {draw = none},
        zticklabels = {},
        xmin = -2,
        xmax = 11,
        ymin = -0.1,
        ymax = 1.1,
        zmin = 0.5,
        zmax = 1.5,
        view={20}{20},
        colormap/blackwhite,
        ]

        \draw[line width = 1pt] (axis cs: -1, 0, \h) to  (axis cs: 0, 0, \h);

        \draw[-{Classical TikZ Rightarrow}] (axis cs: 0,0,\h) to node [above, pos = 0.5] {$p$} (9,0.75+0.2871,\h+0.2871);

        \draw[dashed] (axis cs: 0,0,\h) to (axis cs: 3.2,-0.5334,\h);
        \path (axis cs: 1.92,-0.3192,\h) edge [bend right = 10] (axis cs: 1.9464,0,\h);
        \node at (axis cs: 1.4,-0.13,\h) {$u_{\gamma,i}$};

        \path [draw=violet, fill=none] 
        (axis cs: 0, -0.0043412, \h+0.02462019) --
        (axis cs: 0, 0.3895819, \h+0.09407946) --
        (axis cs: 0, 0.3939231,\h+0.06945927) --
        (axis cs: 0, 0.11150439, \h-0.05649576) --
        cycle;
        \path [draw=violet, fill=none] 
        (axis cs: 0, -0.01915111, \h-0.01606969) --
        (axis cs: 0, -0.27626615, \h+0.29034809) --
        (axis cs: 0, -0.25711504,\h+0.30641778) --
        (axis cs: 0, -0.00682543, \h+0.12481352) -- cycle;
        \path [draw=violet, fill=none] 
        (axis cs: 0, 0.02349232, \h-0.0085505) --
        (axis cs: 0, -0.11331574, \h-0.38442755) --
        (axis cs: 0, -0.13680806 ,\h -0.37587705) --
        (axis cs: 0, -0.10467896, \h-0.06831775) -- cycle;
        \addplot3[draw=violet, domain=0:360, variable=t, smooth, fill=white] (0,{0+0.04*cos(t)},{\h+0.04*sin(t)}) -- cycle;
        \addplot3[draw=violet, domain=0:360, variable=t, smooth, fill=white] (0,{0+0.015*cos(t)},{\h+0.015*sin(t)}) -- cycle; 
        \path [draw=none, fill=white]
            (axis cs: 0, 0.04,\h-0.6)  --
            (axis cs: 0, .015,\h) --
            (axis cs: 0, -.015,\h) --
            (axis cs: 0, -0.04,\h-0.6);
        \path [draw=violet, fill=white] 
            (axis cs: 0, 0.05,\h-0.5)  --
            (axis cs: 0, .015,\h)
            (axis cs: 0, -.015,\h) --
            (axis cs: 0, -0.05,\h-0.5);
        \addplot3[draw=violet, thin, domain=0:360, variable=t, smooth, fill=none] (0,{0+0.406*cos(t)},{\h+0.406*sin(t)}) -- cycle;

        \draw[line width = 1pt] (axis cs: -1+9, 0+0.75, \h) to  (axis cs: 0+9, 0+0.75, \h);

        \draw[dashed] (axis cs: 9,0.75,\h) to (axis cs: 9+3.2,0.75-0.5334,\h);
        \path (axis cs: 9+1.92,0.75-0.3192,\h) edge [bend right = 10] (axis cs: 9+1.9464,0.75,\h);
        \node at (axis cs: 9+1.4,0.75-0.13,\h) {$u_{\gamma,j}$};

        \path [draw=violet, fill=none] 
        (axis cs: 9, -0.0043412+0.75, \h+0.02462019) --
        (axis cs: 9, 0.3895819+0.75, \h+0.09407946) --
        (axis cs: 9, 0.3939231+0.75, \h+0.06945927) --
        (axis cs: 9, 0.11150439+0.75, \h-0.05649576) --
        cycle;
        \path [draw=violet, fill=none] 
        (axis cs: 9, -0.01915111+0.75, \h-0.01606969) --
        (axis cs: 9, -0.27626615+0.75, \h+0.29034809) --
        (axis cs: 9, -0.25711504+0.75, \h+0.30641778) --
        (axis cs: 9, -0.00682543+0.75, \h+0.12481352) -- cycle;
        \path [draw=violet, fill=none] 
        (axis cs: 9, 0.02349232+0.75, \h-0.0085505) --
        (axis cs: 9, -0.11331574+0.75, \h-0.38442755) --
        (axis cs: 9, -0.13680806+0.75, \h -0.37587705) --
        (axis cs: 9, -0.10467896+0.75, \h-0.06831775) -- cycle;
        \addplot3[draw=violet, domain=0:360, variable=t, smooth, fill=white] (9,{0+0.75+0.04*cos(t)},{\h+0.04*sin(t)}) -- cycle;
        \addplot3[draw=violet, domain=0:360, variable=t, smooth, fill=white] (9,{0+0.75+0.015*cos(t)},{\h+0.015*sin(t)}) -- cycle; 
        \path [draw=none, fill=white]
            (axis cs: 9, 0.04+0.75,\h-0.6)  --
            (axis cs: 9, .015+0.75,\h) --
            (axis cs: 9, -.015+0.75,\h) --
            (axis cs: 9, -0.04+0.75,\h-0.6);
        \path [draw=violet, fill=white] 
            (axis cs: 9, 0.05+0.75,\h-0.5)  --
            (axis cs: 9, .015+0.75,\h)
            (axis cs: 9, -.015+0.75,\h) --
            (axis cs: 9, -0.05+0.75,\h-0.5);
        \addplot3[draw=violet, thin, domain=0:360, variable=t, smooth, fill=none] (9,{0+0.75+0.406*cos(t)},{\h+0.406*sin(t)}) -- cycle;

        \draw[dashed] (axis cs: 9, {0.75-0.406}, {\h-0.406}) -- (axis cs: 9, {0.75+0.406}, {\h-0.406}) -- (axis cs: 9, {0.75+0.406}, {\h+0.406}) -- (axis cs: 9, {0.75-0.406}, {\h+0.406}) -- cycle;

        \draw[line width = 1pt, ->] (axis cs: 0+9, 0+0.75, \h) to  (axis cs: 2.5+9, 0+0.75, \h) node [right] {$x_{j}$};
        \draw[line width = 1pt, ->] (axis cs: 0+9, -0.5+0.75, \h) to (axis cs: 0+9, 0.5+0.75, \h) node [right] {$y_{j}$};
        \draw[line width = 1pt, ->] (axis cs: 0+9, 0+0.75, \h-0.4) to (axis cs: 0+9, 0+0.75, 1.5) node [left, anchor=north east] {$z_{j}$};

        \draw[line width = 1pt, ->] (axis cs: 0, 0, \h) to  (axis cs: 2.5, 0, \h) node [right] {$x_i$};
        \draw[line width = 1pt, ->] (axis cs: 0, -0.5, \h) to  (axis cs: 0, 0.5, \h) node [right] {$y_i$};
        \draw[line width = 1pt, ->] (axis cs: 0, 0, \h-0.4) to (axis cs: 0, 0, 1.5) node [left, anchor=north east] {$z_i$};

    \end{axis}
\end{tikzpicture}

%% file: mpc.tex

\section{Model Predictive Control}\label{sec:mpc}

Using the model from \cref{sec:model}, the behaviour of the wind farm can be predicted over a finite horizon based on the current state of the system.
In what follows, predicted signals are denoted with the time index ${(n|k)}$, where $n\in[k,k+N_p]$ with ${N_p\in\N}$ being the controller's prediction horizon.
In this notation, $k$ marks the current discrete time instant, whereas $n$ denotes the prediction time instant.
\begin{assumption}\label{ass:availableStates}
    The states in \eqref{eq:dynamics} are available at time $k$ such that ${\gamma_i(k|k) = \gamma_i(k)}$ and ${a_i(k|k) = a_i(k)}$.
    This can be achieved using, e.g., the FLORIDyn based concept presented in \cite{Becker_2022Obsv}.
\end{assumption}

For $u_{a,i}$, we demand
\begin{subequations}\label{eq:bounds}
    \begin{equation}\label{eq:boundsA}
        0.06 \leq u_{a,i}(n|k)\leq 0.33
    \end{equation}
    for all $i\in[1,N_T]$.
    These bounds are chosen for two reasons.
    $\left(1\right)$ the blade element momentum theory \cite{Hansen_2015} is only valid for $0< u_{a,i}\leq 0.4$.
    $\left(2\right)$ the power coefficient $C_{p,i}(u_{a,i})$ is at its maximum value for $u_{a,i}=0.33$.
    We can reduce $C_{p,i}$ by either increasing $u_{a,i}$ above $0.33$ or by decreasing it.
    Since $C_t$ is related to the thrust force acting on the rotor, which again is associated with mechanical stress, lower values for $u_{a,i}$ are preferable as they lead to lower values of $C_t$.
    The lower bound is chosen as a margin to obtain solutions that keep $C_t>0$.

    Based on \cref{ass:yaw}, we introduce the box constraint
    \begin{equation}\label{eq:boundsGamma}
        -\nicefrac{\pi}{6}\leq u_{\gamma,i}(n|k)\leq\nicefrac{\pi}{6}
    \end{equation}
    for all $i\in[1,N_T]$.
\end{subequations}
Additionally, we have the constraint
\begin{equation}\label{eq:boundsRateGamma}
    -\Delta u_\gamma \leq u_{\gamma,i}(n|k)-u_{\gamma,i}(n-1|k) \leq \Delta u_\gamma
\end{equation}
for all $i\in[1,N_T]$ to account for the limited yaw actuation speed.
Note, that we do not implement an analogous constraint for $u_{a,i}$ since this part of the \ac{wt} dynamics is expected to be negligible (recall \cref{ass:dynamics}).
During power tracking, we expect lower values of $u_{a,i}$ for which $x_c$ in \eqref{eq:x0} becomes larger.
To uphold \cref{ass:distance}, we employ
\begin{equation}\label{eq:boundsXc}
    x_c(n|k) \leq \leftindex^i x_j
\end{equation}
for all $j\in[1,N_T]$ and $i\in\mathbb{U}_j$ so that we have far wake conditions at downstream \acp{wt}.
Finally, we want to limit all power outputs by the \acp{wt}' rated power $P_r\in\Rp$, i.e.,
\begin{equation}\label{eq:boundsPower}
    P_i(n|k) \leq P_r.
\end{equation}

We choose the cost function
\begin{equation}\label{eq:cost}
    \ell(k) = \textstyle\sum_{n=k}^{k+N_p} \left(q_P\left(P_F(n|k)-P_{ref}(k)\right)^2 +q_{P,2}\textstyle\sum_i^{N_T}P_i(n|k)^2\right)
\end{equation}
with weightings ${q_P,q_{P,2}\in\Rps}$.
The cost function aims to track the reference power $P_{ref}(k)$ and to distribute the farm power evenly among all \acp{wt}.
Let us define the control matrix
\begin{equation}
    U_i(k) = 
    \begin{bmatrix}
        u_{\gamma,i}(k|k) & u_{\gamma,i}(k+1|k) & \cdots & u_{\gamma,i}(k+N_p|k)\\
        u_{a,i}(k|k) & u_{a,i}(k+1|k) & \cdots & u_{a,i}(k+N_p|k)
    \end{bmatrix}
\end{equation}
and form $U(k)=[U_1(k),\cdots,U_{N_T}(k)]$ to state the following nonlinear optimization problem.
\begin{problem}\label{problem:nonlinear}
    \begin{subequations}
        \begin{equation*}
            \min_{U(k)}\quad\ell\left(k\right)
        \end{equation*}
        subject to
        \begin{align*}
            \begin{bmatrix}
                \gamma_i(n+1|k)\\
                a_i(n+1|k)
            \end{bmatrix}
            =
            \begin{bmatrix}
                A & 0\\
                0 & A
            \end{bmatrix}
            \begin{bmatrix}
                \gamma_i(n|k)\\
                a_i(n|k)
            \end{bmatrix}
            +
            \begin{bmatrix}
                B & 0\\
                0 & B
            \end{bmatrix}
            \begin{bmatrix}
                u_{\gamma,i}(n|k)\\
                u_{a,i}(n|k)
            \end{bmatrix}
        \end{align*}
    \end{subequations}
    as well as \labelcref{eq:thrustCoeffcient,eq:x0,eq:sigma,eq:C,eq:zeta,eq:deltay,eq:rInt,eq:rotorEffectiveWindSpeed,eq:powerCoefficient,eq:turbinePower,eq:farmPower,eq:bounds,eq:boundsRateGamma,eq:boundsXc,eq:boundsPower} $\forall n\in[k,k+N_p]$ and $i\in[1,N_T]$ with given initial conditions $\gamma_i(k|k)=\gamma_i(k)$, $a_i(k|k)=a_i(k)$ and $u_{\gamma,i}(k-1|k)=u_{\gamma,i}(k-1)$ which refers to the previous yaw misalignment.
    
    The solutions obtained for $u_{\gamma,i}(k|k)$ and $u_{a,i}(k|k)$ are used to actuate the wind farm.
    At $k+1$, new initial values are available and the process is repeated to account for disturbances and model inaccuracies which gives us a moving horizon control scheme.
\end{problem}
\Cref{problem:nonlinear} is a nonlinear optimization problem.
While nonlinear solvers are available, they may perform poorly in terms of solving times and may not find a globally optimal solution.
Since we are interested in real-time control, we will reformulate \cref{problem:nonlinear} as an \ac{miqcqp} for which the global optimum can be found by commercial solvers in reasonable time.

Constraint \eqref{eq:boundsGamma} allows us to employ the second order Taylor approximations

\vspace{3mm}
\begin{minipage}{0.4\textwidth}
    \noindent
    \setcounter{equation}{22}
    \begin{equation}\label{eq:approxCos}
        \cos\left(u_{\gamma,i}\right) \approx \textstyle 1-\frac{u_{\gamma,i}^2}{2}
    \end{equation}
\end{minipage}
\quad and
\begin{minipage}{0.4\textwidth}
    \noindent
    \begin{equation}\label{eq:approx1_Cos}
        \textstyle\frac{1}{\cos(u_{\gamma,i})} \approx \textstyle 1+\frac{u_{\gamma,i}^2}{2}.
    \end{equation}
\end{minipage}
\vspace{3mm}

The wake core length $x_c$ \eqref{eq:x0}, wake centre's wind speed deficit $r_C$ \eqref{eq:C} and wake centre's spanwise deflection $\delta_y$ \eqref{eq:deltay} are approximated by polynomials in $\gamma$ and $a$.
This yields
\begin{equation}\label{eq:approxX0}
    x_c(\gamma,a) = K_{20}^{x_c}\gamma^2 + K_{02}^{x_c}a^2 + K_{01}^{x_c}a + K_{00}^{x_c},
\end{equation}
where ${K_{00}^{x_c},K_{01}^{x_c},K_{02}^{x_c},K_{20}^{x_c}}\in\R$ are precomputed based on wind conditions.
For $r_C$ and $\delta_y$, the resulting polynomials are
\begin{align}
    r_C(\gamma,a) &= K_{03}^C a^3 + K_{21}^C\gamma^2 a + K_{20}^C\gamma^2 + K_{02}^Ca^2 + K_{01}^C a + K_{00}^C\label{eq:approxC}\\
    \delta_y(\gamma,a) &= K_{30}^{\delta_y}\gamma^3 + K_{12}^{\delta_y}\gamma a^2 + K_{11}^{\delta_y}\gamma a + K_{10}^{\delta_y}\gamma,\label{eq:approxDY}
\end{align}
where the polynomial coefficients, e.g., $K_{03}^C$, $K_{00}^{\delta_y}$, depend on the farm's geometry and are computed offline yielding approximation errors below $\SI{17}{\percent}$ for turbine distances within $[5D,15D]$.
To reformulate the remaining equations, auxiliary variables and additional equality constraints are introduced which are displayed by \cref{tab:auxiliary}.

\begin{table}[b]
    \centering
    \caption{auxiliary variables}
    \label{tab:auxiliary}
    \scalebox{0.7}{
        \begin{tabular}{ccccc}
            \toprule
                        & Quadratic     & &               & Quadratic \\
            Original term & reformulation & & Original term & reformulation \\
            \cmidrule{1-2} \cmidrule{4-5} 
            $\cos(u_{\gamma})$ & $\xi_1=1-\frac{u_\gamma}{2}$ && 
            $\frac{\leftindex^i y_j+\frac{1}{2}D\cos(u_{\gamma,j}-\delta_y)}{\sqrt{2}\sigma_y}$ & $\xi_{11}\sqrt{2}\sigma_y = \leftindex^i y_j+\frac{1}{2}D\xi_1-\delta_y$\\
            \cmidrule{1-2} \cmidrule{4-5} 
            $\gamma^2$ & $\xi_2=\gamma^2$ &&
            $\frac{\leftindex^i y_j-\frac{1}{2}D\cos(u_{\gamma,j}-\delta_y)}{\sqrt{2}\sigma_y}$ & $\xi_{12}\sqrt{2}\sigma_y = \leftindex^i y_j-\frac{1}{2}D\xi_1-\delta_y$\\
            \cmidrule{1-2} \cmidrule{4-5} 
            $\gamma^3$ & $\xi_3 = \xi_2\gamma$ &&
            $\frac{D}{\sqrt{8}\sigma_z}$ & $\xi_{13}\sqrt{8}\sigma_z = D$\\
            \cmidrule{1-2} \cmidrule{4-5} 
            $a^2$ & $\xi_4 = a^2$ &&
            $r_C\sigma_y$ & $\xi_{14} = r_C\sigma_y$\\
            \cmidrule{1-2} \cmidrule{4-5} 
            $a^3$ & $\xi_5 = \xi_4 a$ &&
            $r_C\sigma_y\sigma_z$ & $\xi_{15} = \xi_{14}\sigma_z$\\
            \cmidrule{1-2} \cmidrule{4-5} 
            $(1-u_a)^2$ & $\xi_6 = (1-u_a)^2$ &&
            $\frac{\pi r_C\sigma_y\sigma_z}{D^2\cos(u_\gamma)}$ & $\xi_{16}D^2 = \xi_{15}\pi(2-\xi_1)$\\
            \cmidrule{1-2} \cmidrule{4-5} 
            $C_p$ & $\xi_7 = \kappa 4u_a\xi_6$ &&
            $\frac{\pi r_C\sigma_y\sigma_z}{D^2\cos(u_\gamma)}\erf\left(\frac{D}{\sqrt{8}\sigma_z}\right)$ & $\xi_{17} = \xi_{16}\erfpwa(\xi_{13})$\\
            \cmidrule{1-2} \cmidrule{4-5} 
            $V^2$ & $\xi_8 = V^2$ &&
            $r_C$ & $\xi_{18}\left(\erfpwa(\xi_{11})-\erfpwa(\xi_{12})\right)$\\
            \cmidrule{1-2} \cmidrule{4-5} 
            $V^3$ & $\xi_9 = \xi_8V$ && $\cos(u_\gamma)^{p_p}$ & $\xi_{19} = \xi_1^2$\\
            \cmidrule{1-2} \cmidrule{4-5} 
            $C_pV^3$ & $\xi_{10} = \xi_{7}\xi_{9}$ && $P$ & $\xi_{20} = \frac{\eta\rho\pi D^2}{8}\xi_{18}\xi_{19}$\\
            \bottomrule
        \end{tabular}
    }
\end{table}

Finally, the function $\erf(\xi)$ is approximated by a piecewise affine function $\erfpwa(\xi)$ (see also \cite{Bemporad_1999,Sterle2023}).
These changes allow us to reformulate \cref{problem:nonlinear} as the following \ac{miqcqp}.
\begin{problem}\label{problem:quadratic}
    \begin{subequations}
        \begin{equation*}
            \min_{U(k)}\quad\ell\left(k\right)
        \end{equation*}
        subject to
        \begin{align*}
            \begin{bmatrix}
                \gamma_i(n+1|k)\\
                a_i(n+1|k)
            \end{bmatrix}
            =
            \begin{bmatrix}
                A & 0\\
                0 & A
            \end{bmatrix}
            \begin{bmatrix}
                \gamma_i(n|k)\\
                a_i(n|k)
            \end{bmatrix}
            +
            \begin{bmatrix}
                B & 0\\
                0 & B
            \end{bmatrix}
            \begin{bmatrix}
                u_{\gamma,i}(n|k)\\
                u_{a,i}(n|k)
            \end{bmatrix}
        \end{align*}
    \end{subequations}
    as well as \labelcref{eq:farmPower,eq:bounds,eq:boundsRateGamma,eq:boundsXc,eq:boundsPower,eq:approxCos,eq:approx1_Cos,eq:approxX0,eq:approxC,eq:approxDY} and \cref{tab:auxiliary} ${\forall n\in[k,k+N_p]}$ and ${i\in[1,N_T]}$ with given initial conditions $\gamma_i(k|k)=\gamma_i(k)$, $a_i(k|k)=a_i(k)$ and $u_{\gamma,i}(k-1|k)=u_{\gamma,i}(k-1)$.
\end{problem}

%% file: caseStudy.tex

\section{Case Study}
\label{sec:caseStudy}

\begin{table}[b]
    \centering
    \caption{farm and controller parameters}
    \label{tab:turbine}
    \scalebox{0.7}{
        \begin{tabular}{ccccccccccc}
            \toprule
            Parameter & Value &&
            Parameter & Value &&
            Parameter & Value &&
            Parameter & Value\\
            \cmidrule{1-2}
            \cmidrule{4-5}
            \cmidrule{7-8} 
            \cmidrule{10-11} 
            $T_s$ & $\SI{13}{\second}$
            &&
            $V_\infty$ & $\SI{10}{\meter\per\second}$
            &&
            $D$ & $\SI{130}{\meter}$
            &&
            $\rho$ & $\SI{1.225}{\kilogram\per\meter^3}$
            \\
            \cmidrule{1-2} \cmidrule{4-5} \cmidrule{7-8} \cmidrule{10-11} 
            $I$ & $\SI{6}{\percent}$
            &&
            $k_y,k_z$ & $0.0267$
            &&
            $\eta$ & $0.9367$
            &&
            $P_r$ & $\SI{3.35}{\mega\watt}$
            \\
            \cmidrule{1-2} \cmidrule{4-5} \cmidrule{7-8} \cmidrule{10-11} 
            $N_p$ & $8$
            &&
            $\kappa$ & $0.8174$
            &&
            $\Delta u_\gamma$ & $0.0572$
            &&
            $\leftindex^1y_3$ & $-0.75D$
            \\
            \cmidrule{1-2} \cmidrule{4-5} \cmidrule{7-8} \cmidrule{10-11} 
            $\leftindex^2y_3$ & $0.75D$
            &&
            $q_P$ & $10^{-8}\si{\per\watt}$
            &&
            $\leftindex^1x_3,\leftindex^2x_3$ & $7D$
            &&
            $q_{P,2}$ & $10^{-12}\si{\per\watt}$\\
            \bottomrule
        \end{tabular}
    }
\end{table}

In what follows, \cref{problem:quadratic} will be used to operate the farm in \cref{fig:threeTurbineFarm}. 
The \ac{mpc} and farm parameters, which are based on IEA $\SI{3.4}{\mega\watt}$ turbines \cite{RWT} are displayed in \cref{tab:turbine}.
The prediction horizon $N_p$ is chosen such that the effects of $u_{\gamma,1}$ and $u_{\gamma,2}$ on \ac{wt} $3$ are considered within the prediction.
The \ac{mpc} is implemented in MATLAB using Yalmip \cite{Lofberg2004} and Gurobi.
All simulations are conducted on a MacBook Air with an Apple M1 chip with 8 cores and $\SI{8}{\mathrm{GB}}$ RAM.
The wind farm is simulated using FLORIDyn .

\begin{figure}
    \centering
    \hspace{-3cm}
    \vspace{-1cm}
    \subcaptionbox*{}{\input{figures/outputLegend}}\\
    \vspace{-0.3cm}
    \subcaptionbox*{}{\input{figures/outputPower}}
    \subcaptionbox*{}{\input{figures/outputTurbine}}\\
    \vspace{-0.5cm}
    \subcaptionbox*{}{\input{figures/outputGamma}}
    \subcaptionbox*{}{\input{figures/outputA}}
    \subcaptionbox*{}{\input{figures/solverTime}}
    \vspace{-0.8cm}
    \caption{
        Simulation results for three turbine farm.
        The results for combined farm power, turbine power, yaw misalignment, axial induction factor and solver times are depicted.
        The controller's sampling time is indicated by a dashed black line in (e).
        }
    \label{fig:results}
\end{figure}
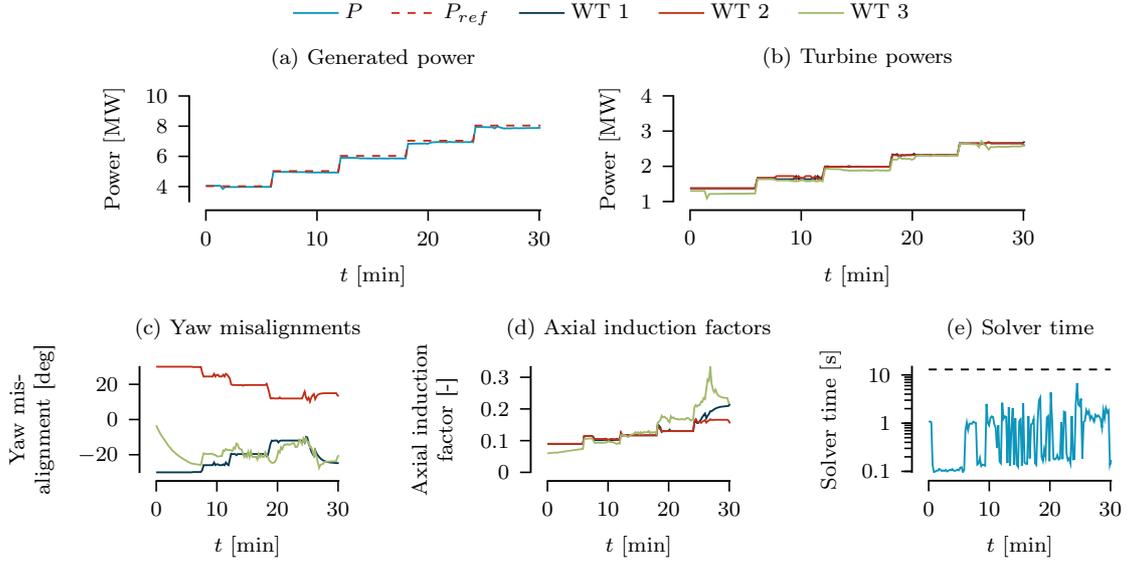

The results in \cref{fig:results} indicate that the farm approximately tracks the reference power while evenly distributing infeed among all \acp{wt}, which indicates that mechanical loads are also well distributed.
To achieve lower power outputs, the upstream \acp{wt}' yaw misalignments are set such that wakes are steered towards \ac{wt} $3$ and its available power is reduced.
With increasing reference power the yaw misalignments are updated such that the wakes are steered away from \ac{wt} $3$ to increase its available power.
Thus, higher power outputs can be achieved.
The axial induction factors are not evenly distributed among all turbines. 
While those of \ac{wt} $1$ and $2$ appear similar, the one of \ac{wt} $3$ differs to compensate wake effects.
Its noticeable that the yaw misalignment of \ac{wt} $3$ is further used to control its power output as no other turbine is affected by its wake.
Finally, the solver time in \cref{fig:results} (e) remains below the sampling time during the simulation which indicates real-time applicability of the \ac{mpc}.

%% file: figures/outputLegend.tex

\tikzset{external/export next=false}
  \begin{tikzpicture}[font=\scriptsize]
    \begin{axis}[
      myPlot,
      height = 1.7cm,
      clip = true,
      x axis line style = {draw = none},
      x tick style = {draw = none},
      xticklabels = {},
      y axis line style = {draw = none},
      y tick style = {draw = none},
      yticklabels = {},
      legend columns=5,
      legend style={
        at={(1.8,1.4)},
        yshift = .2,
        anchor=south east,
        draw=none,
        fill=none,
        legend cell align=left,
        /tikz/every even column/.append style={column sep=0.3cm},
        legend entries = {$P$, $P_{ref}$, \ac{wt} 1, \ac{wt} 2, \ac{wt} 3},
        },
      ]
      
    \addplot[vgLightBlue, line width = 0.7pt] coordinates {
      (0,0)
    };
    \addplot[dashed, vgRed, line width = 0.7pt] coordinates {
      (1,1)
    };
    \addplot[vgDarkBlue, line width = 0.7pt] coordinates {
        (0,0)
      };
    \addplot[vgRed, line width = 0.7pt] coordinates {
        (0,0)
      };
    \addplot[vgGreen, line width = 0.7pt] coordinates {
        (0,0)
      };
  \end{axis}
\end{tikzpicture}

%% file: figures/outputPower.tex

\tikzset{external/export next=false}
  \begin{tikzpicture}[font=\scriptsize]
    \begin{axis}[
      myPlot,
      width = 6cm,
      height = 3cm,
      xlabel = {$t$ $[\si{\min}]$},
      ylabel = {Power $[\si{\mega\watt}]$},
      ymax = 10,
      ymin = 3,
      title = {(a) Generated power},
      clip = true,
      ]
      
    \addplot[vgLightBlue, line width = 0.7pt] table [x = {t}, y = {P}, col sep=comma] {figures/csvFiles/caseStudy.csv};
    \addplot[dashed,vgRed, line width = 0.7pt] table [x = {t}, y = {Pr}, col sep=comma] {figures/csvFiles/caseStudy.csv};
  \end{axis}
\end{tikzpicture}

%% file: figures/outputTurbine.tex

\tikzset{external/export next=false}
  \begin{tikzpicture}[font=\scriptsize]
    \begin{axis}[
      myPlot,
      width = 6cm,
      height = 3cm,
      xlabel = {$t$ $[\si{\min}]$},
      ylabel = {Power $[\si{\mega\watt}]$},
      ymax = 4,
      ymin = 1,
      title = {(b) Turbine powers},
      clip = true,
      ]
      
    \addplot[vgLightBlue, line width = 0.7pt] coordinates {
      (0,0)
    };
    \addplot[dashed, vgRed, line width = 0.7pt] coordinates {
      (0,0)
    };
    \addplot[vgDarkBlue, line width = 0.7pt] table [x = {t}, y = {P1}, col sep=comma] {figures/csvFiles/caseStudy.csv};
    \addplot[vgRed, line width = 0.7pt] table [x = {t}, y = {P2}, col sep=comma] {figures/csvFiles/caseStudy.csv};
    \addplot[vgGreen, line width = 0.7pt] table [x = {t}, y = {P3}, col sep=comma] {figures/csvFiles/caseStudy.csv};
  \end{axis}
\end{tikzpicture}

%% file: figures/outputGamma.tex

\tikzset{external/export next=false}
  \begin{tikzpicture}[font=\scriptsize]
    \begin{axis}[
      myPlot,
      width = 4cm,
      height = 3cm,
      xlabel = {$t$ $[\si{\min}]$},
      ylabel = {\begin{tabular}{c}Yaw mis- \\ alignment [deg]\end{tabular}},
      ymax = 30,
      ymin = -30,
      title = {(c) Yaw misalignments},
      clip = true,
      ]
      
    \addplot[vgLightBlue, line width = 0.7pt] coordinates {
      (0,0)
    };
    \addplot[dashed, vgRed, line width = 0.7pt] coordinates {
      (0,0)
    };
    \addplot[vgDarkBlue, line width = 0.7pt] table [x = {t}, y expr = -\thisrow{gamma1}, col sep=comma] {figures/csvFiles/caseStudy.csv};
    \addplot[vgRed, line width = 0.7pt] table [x = {t}, y expr = -\thisrow{gamma2}, col sep=comma] {figures/csvFiles/caseStudy.csv};
    \addplot[vgGreen, line width = 0.7pt] table [x = {t}, y expr = -\thisrow{gamma3}, col sep=comma] {figures/csvFiles/caseStudy.csv};
  \end{axis}
\end{tikzpicture}

%% file: figures/outputA.tex

\tikzset{external/export next=false}
  \begin{tikzpicture}[font=\scriptsize]
    \begin{axis}[
      myPlot,
      width = 4cm,
      height = 3cm,
      xlabel = {$t$  $[\si{\min}]$},
      ylabel = {\begin{tabular}{c}Axial induction \\ factor [-]\end{tabular}},
      ymax = 0.3333,
      ymin = 0,
      title = {(d) Axial induction factors},
      clip = true,
      ]
      
    \addplot[vgDarkBlue, line width = 0.7pt] table [x = {t}, y = {a1}, col sep=comma] {figures/csvFiles/caseStudy.csv};
    \addplot[vgRed, line width = 0.7pt] table [x = {t}, y = {a2}, col sep=comma] {figures/csvFiles/caseStudy.csv};
    \addplot[vgGreen, line width = 0.7pt] table [x = {t}, y = {a3}, col sep=comma] {figures/csvFiles/caseStudy.csv};
  \end{axis}
\end{tikzpicture}

%% file: figures/solverTime.tex

\tikzset{external/export next=false}
  \begin{tikzpicture}[font=\scriptsize]
    \begin{axis}[
      myPlot,
      width = 4cm,
      height = 3cm,
      ymode = log,
      title = (e) Solver time,
      xlabel = {$t$ $[\si{\minute}]$},
      ylabel = {Solver time $[\si{\second}]$},
      clip = true,
      ytick = {0.1,1,10},
      yticklabels = {0.1,1,10},
      ymax = 15,
      ]
      
    \addplot[vgLightBlue] table [x = {t}, y = {tSolve}, col sep=comma] {figures/csvFiles/caseStudy.csv};

    \draw[dashed] (axis cs:0,13) -- (axis cs:30,13);
  \end{axis}
\end{tikzpicture}

%% file: conclusions.tex

\section{Conclusions}\label{sec:conclusion}

In this paper, a mixed-integer quadratically-constrained quadratic model predictive wind farm controller was presented.
Our model based approach considers time-dependent wake dynamics.
We could show that our approach is capable of approximately tracking a given reference power over a large range by combining induction control and yaw misaligning.
Additionally, the power output is evenly distributed between each turbine.

The presented controller provides a basis for wake modelling receding horizon control schemes.
Further research shall investigate options to keep the mechanical loads evenly distributed and lift some assumptions such as static wind conditions and the far wake limitation (\cref{ass:wind,ass:wakeVelocity,ass:turbulence,ass:distance}).
Furthermore, the controller shall be tested on more complex wind farm topologies and in high-fidelity simulations such as SOWFA to further validate its real world applicability.
Finally, the wind farm controller shall be employed with individual \acl{wt} controllers, e.g., from \cite{Grapentin2022}.